\documentclass[twocolumn,aps,prd,notitlepage,showpacs,nofootinbib,superscriptaddress]{revtex4-1}

\usepackage{graphicx}
\usepackage[utf8]{inputenc} 

\usepackage{amsmath}
\usepackage{amssymb}
\usepackage{float}
\usepackage{comment}
\usepackage{slashed}
\usepackage[normalem]{ulem}
\usepackage{physics}

\usepackage{booktabs}
\usepackage{mathtools,braket,blkarray}

\usepackage[usenames,dvipsnames]{color}
\usepackage[colorinlistoftodos]{todonotes}
\usepackage[colorlinks=true,citecolor=darkred,urlcolor=darkred, pdfborder={0 0 0}]{hyperref}
\usepackage[normalem]{ulem}

\usepackage{multirow}
\definecolor{darkred}{rgb}{0.6,0,0}
\definecolor{brown}{rgb}{0.59, 0.29, 0.0}

\usepackage[colorinlistoftodos]{todonotes}

\definecolor{linkcolor}{rgb}{0,0,0.5}

\usepackage{soul}

\begin{document}
\bibliographystyle{unsrt}

\title{\boldmath \color{BrickRed}  Quantization of second order fermions}

\author{Rodolfo Ferro-Hern\'andez}\email{rferrohe@uni-mainz.de}
\affiliation{PRISMA+ Cluster of Excellence and Institute for Nuclear Physics Johannes Gutenberg University, 55099 Mainz, Germany.}

\author{Julio Olmos}\email{jc.olmosgomez@ugto.mx}
\affiliation{Departamento de F\'isica, DCI, Campus Le\'on, Universidad de
  Guanajuato, Loma del Bosque 103, Lomas del Campestre C.P. 37150, Le\'on, Guanajuato, M\'exico}

\author{Eduardo Peinado}\email{epeinado@fisica.unam.mx}
\affiliation{Instituto de F\'isica, Universidad Nacional Aut\'onoma de M\'exico,
A.P. 20-364, Ciudad de M\'exico 01000, Mexico}
\affiliation{Departamento de F\'isica, Centro de Investigaci\'on y de Estudios Avanzados del Instituto Polit\'ecnico Nacional\\
Apartado Postal 14-740, 07000 Ciudad de M\'exico, Mexico}
\author{Carlos A. Vaquera-Araujo}\email{vaquera@fisica.ugto.mx}
\affiliation{Consejo Nacional de Humanidades, Ciencias y Tecnolog\'ias, Av. Insurgentes Sur 1582. Colonia Cr\'edito Constructor, Del. Benito Ju\'arez, C.P. 03940, Ciudad de M\'exico, M\'exico}
\affiliation{Departamento de F\'isica, DCI, Campus Le\'on, Universidad de
  Guanajuato, Loma del Bosque 103, Lomas del Campestre C.P. 37150, Le\'on, Guanajuato, M\'exico}
\affiliation{Dual CP Institute of High Energy Physics, C.P. 28045, Colima, Mexico}


\begin{abstract}
\vspace{0.5cm}

The quantization of a massive spin-$1/2$ field that satisfies the Klein-Gordon equation is studied. The framework is consistent, provided it is formulated as a pseudo-Hermitian quantum field theory by the redefinition of the dual field and the identification of an operator that modifies the inner product of states in Hilbert space to preserve a real energy spectrum and unitary evolution. Since the fermion field has mass dimension one, the theory admits renormalizable fermion self-interactions.

\end{abstract}

\maketitle
\noindent

In this paper, we introduce a consistent pseudo-Hermitian quantum field theory for second-order massive fermions $\psi$ transforming under the $(\tfrac{1}{2},0)\oplus(0,\tfrac{1}{2})$ representation of the restricted Lorentz group (RLG). LeClair and Neubert \cite{LeClair:2007iy} presented the first successful second order formalism describing the dynamics of $N$-component complex fermions in a pseudo-Hermitian framework. In that work, the authors studied two cases: 1) If the fields transform as scalars in the $(0,0)$ representation of the RLG, the theory is automatically Lorentz invariant and the usual spin-statistics connection is evaded, yielding a free theory of scalar fermions obeying the Klein-Gordon equation; 2) Insisting that the fields are spin-$1/2$ fermions and therefore preserving the usual spin-statistics connection, there is a subgroup $\mathrm{SO}(3)$ of the full global symplectic $\mathrm{Sp}(2N)$ group, displayed by the theory, that can be identified with the rotational symmetry that defines spin-$1/2$ particles in a nonrelativistic setup. However, due to the Lorentz group not being a subgroup of $\mathrm{Sp}(2N)$, the model in \cite{LeClair:2007iy} fails to manifest full Lorentz invariance. Here, we build an authentic Lorentz invariant theory for spin-$1/2$ fermions with a kinetic term that is second order in derivatives with real energy spectrum and unitary time evolution. 

A pseudo-Hermitian Lagrangian satisfies 
\begin{equation}
\mathcal{L}^{\#}\equiv \eta^{-1}\mathcal{L}^\dagger\eta=\mathcal{L}.
\label{pseudo-Hermitian}
\end{equation}
This generalization of Hermiticity was proposed in \cite{Mostafazadeh:2001jk, Mostafazadeh:2008pw} extending the results of $PT$-symmetric quantum mechanics introduced in \cite{Bender:1998ke}, and has two important features; 1) the energy spectrum of the theory is real and 2) time evolution is unitary upon the definition of a suitable internal product of states with the aid of the $\eta$ operator.

The Lagrangian for the $\psi$ field is given by
\label{sec:formalism}
\begin{equation}
\mathcal{L}= \partial^{\mu}\widehat{\psi}\partial_\mu\psi-m^2\widehat{\psi}\psi ,
\label{Lag1}
\end{equation}
where $\widehat{\psi}$ is not the Dirac adjoint of $\psi$, but instead a redefinition of its dual that renders the theory pseudo-Hermitian. This redefinition also eliminates the presence of negative norm states in the spectrum \cite{KP}, which appear when the standard Dirac adjoint is used in this Lagrangian. 

The first observation we can make about this theory is that the conjugate momenta are $\pi_\psi=\dot{\widehat\psi}$, $\pi_{\widehat{\psi}}=\dot{\psi}$, where the dot denotes the derivative with respect to time. Without imposing further constraints, this formalism involves a spinor with four complex (eight real) components, and the phase space is parametrized by both $\psi$ and $\dot{\widehat{\psi}}$, yielding 16 real components. Thus, the second-order formalism describes a field with eight degrees of freedom. In contrast, the momentum conjugate associated to a four component Dirac spinor $\chi$ is $i\chi^\dagger$, and therefore, the whole phase space is parametrized by the eight real components of $\chi$, giving a total number of four real degrees of freedom for the Dirac field.
A second observation about the second-order formalism is that the field $\psi$ has mass dimension one, in sharp contrast with Dirac spinors, which have mass dimension $3/2$.

It can be shown that the most general solution to the Klein-Gordon equation for spin-$1/2$ fields can be written in terms of two solutions to the Dirac equation $\chi_{1}$ and $\chi_2$ of the form \cite{CufaroPetroni:1985tu}
\begin{equation}
\psi=\frac{1}{\sqrt{2m}}(\chi_1+\gamma^5\chi_2).
\end{equation}
The plane wave expansions of $\psi$ and $\widehat\psi$ are
\begin{widetext}
\begin{align}
\psi(x)=&\int \frac{d^3\mathbf{p}}{(2\pi)^{3}2\sqrt{m\omega_{\mathbf{p}}}}\sum_s \bigg\{\left[u^{s}_{\mathbf{p}}a^{1 s}_{\mathbf{p}}+\gamma^5u^{s}_{\mathbf{p}}a^{2 s}_{\mathbf{p}}\right]e^{-ip\cdot x}+\left[v^{s}_{\mathbf{p}}b^{1 s\dagger}_{\mathbf{p}}+\gamma^5v^{s}_{\mathbf{p}}b^{2 s\dagger}_{\mathbf{p}}\right]e^{ip\cdot x}\bigg\},\nonumber\\
\widehat{\psi}(x)=&\int \frac{d^3\mathbf{p}}{(2\pi)^{3}2\sqrt{m\omega_{\mathbf{p}}}}\sum_s \bigg\{\left[\bar{u}^{s}_{\mathbf{p}}a^{1 s\dagger}_{\mathbf{p}}+\bar{u}^{s}_{\mathbf{p}}\gamma^5a^{2 s\dagger}_{\mathbf{p}}\right]e^{ip\cdot x}+\left[\bar{v}^{s}_{\mathbf{p}}b^{1 s}_{\mathbf{p}}+\bar{v}^{s}_{\mathbf{p}}\gamma^5b^{2 s}_{\mathbf{p}}\right]e^{-ip\cdot x}\bigg\},\label{field_expansion}
\end{align}
\end{widetext}
with $\omega_{\mathbf{p}}=+\sqrt{|\mathbf{p}|^2+m^2}$, $p^\mu=(\omega_{\mathbf{p}},\mathbf{p})$,  $u^{s}_{\mathbf{p}}$, $v^{s}_{\mathbf{p}}$ as the positive and negative energy solutions of the Dirac free equation, and $s=\pm\frac{1}{2}$. The theory is by construction explicitly Lorentz invariant.

The pseudo-Hermitian nature of the formalism emerges from the introduction of a linear and invertible operator $\eta$ defined by the following relations \cite{LeClair:2007iy}:
\begin{equation}
\begin{array}{cccc}
\eta^{-1} a^{j s}_{\mathbf{p}}\eta=(-1)^{j-1} a^{j s}_{\mathbf{p}},\qquad \eta^{-1} b^{j s\dagger}_{\mathbf{p}}\eta= (-1)^{j-1} b^{j s\dagger}_{\mathbf{p}},
\end{array}
\end{equation}
where $j=1,2$.
Notice that the difference between the new dual $\widehat{\psi}$  and the Dirac adjoint $\overline{\psi}$ is the sign of the modes involving $\gamma^5$ in Eq.~(\ref{field_expansion}). The dual field  in terms of $\eta$, is given by $\widehat\psi=\eta^{-1}\bar\psi\eta$. Assuming that the operator $\eta$ 
is Hermitian implies that $\eta^{-1}(\widehat\psi\psi)^\dagger\eta=\widehat\psi\psi$,
fulfilling the pseudo-Hermiticity of the free Lagrangian in Eq.~(\ref{Lag1}), as prescribed in Eq.~(\ref{pseudo-Hermitian}).
If the following anticommutation relations for the momentum-space field operators hold:
\begin{equation}
\begin{split}
\left\{a^{j s}_{\mathbf{p}},a^{k r\dagger}_{\mathbf{p}'}\right\}&=(2\pi)^3\delta^{j k}\delta^{r s}\delta^{(3)}(\mathbf{p}-\mathbf{p}'),\\
\left\{b^{j s}_{\mathbf{p}},b^{k r\dagger}_{\mathbf{p}'}\right\}&=(2\pi)^3\delta^{j k}\delta^{r s}\delta^{(3)}(\mathbf{p}-\mathbf{p}'),
\end{split}
\label{canonicalrelations}
\end{equation}
with all other anticommutators vanishing, an explicit solution for the operator $\eta$ is given by
\begin{equation}
\eta=\exp\left[i\pi\int\frac{d^3\mathbf{p}}{(2\pi)^{3}}\sum_s \left(a^{2 s\dagger}_{\mathbf{p}}a^{2 s}_{\mathbf{p}}+b^{2 s\dagger}_{\mathbf{p}}b^{2 s}_{\mathbf{p}}\right)\right].
\end{equation}
This operator satisfies $\eta=\eta^{-1}=\eta^\dagger$, meaning that it is simultaneously Hermitian and unitary, yielding $\eta^2=1$.

The dynamics of the pseudo-Hermitian quantum theory with $H^{\#}=\eta^{-1} H^\dagger\eta=H$ is guaranteed to be unitary if the inner product between two states is defined as
$
\braket{a|b}_{\eta}\equiv\bra{a}\eta\ket{b}$.
In this way, probability amplitudes are preserved under time evolution and the energy spectrum is real.
The conjugate momenta satisfy the time canonical anticommutation relations
 \begin{equation}\label{canonicalquantization}
 \begin{split}
\left\{\psi_{\alpha}(\mathbf{x},t),\pi_\psi{}_{\beta}(\mathbf{x}',t)\right\}&=-\left\{\widehat{\psi}_{\alpha}(\mathbf{x},t),\pi_{\widehat{\psi}}{}_{\beta}(\mathbf{x}',t)\right\}\\&=i\delta_{\alpha\beta}\delta^{(3)}(\mathbf{x}-\mathbf{x}').
\end{split}
\end{equation}
Due to the redefinition of the dual implemented by the $\eta$ operator, these anticommutation relations lead to the standard non-vanishing canonical relations for the momentum space operators in Eq.~({\ref{canonicalrelations}}). Additionally, it can be shown that the fields display the correct properties under microcausality. In particular, we have
\begin{equation}
\left\{\psi_{\alpha}(x),\psi_{\beta}(y)\right\}=\left\{\widehat{\psi}_{\alpha}(x),\widehat{\psi}_{\beta}(y)\right\}=0,
\end{equation}
and
\begin{equation}
\begin{split}
&\left\{\psi_{\alpha}(x),\widehat{\psi}_{\beta}(y)\right\}\equiv\Delta(x-y)\delta_{\alpha\beta}\\&\qquad=
\int \frac{d^3\mathbf{p}}{(2\pi)^{3}2 \omega_{\mathbf{p}}}\bigg\{e^{-ip\cdot(x-y)}-e^{ip\cdot(x-y)}\bigg\}\delta_{\alpha\beta},
\end{split}
\end{equation}
where $\Delta(x-y)$ is the well-known Lorentz invariant and causal Schwinger's Green function. The Hamiltonian and the momentum operator are given by
\begin{equation}
\begin{split}
H&=:\int d^3\mathbf{x}\bigg\{\dot{\widehat{\psi}}\dot{\psi}+\nabla\widehat\psi\cdot\nabla\psi+m^2\widehat\psi\psi\bigg\}:,\\
\mathbf{P}&=:-\int d^3\mathbf{x}\bigg\{\dot{\widehat{\psi}}\nabla\psi+\nabla\widehat{\psi}\dot{ \psi}\bigg\}:,
\end{split}
\end{equation}  
where $:{}:$ stands for normal ordering. In terms of the momentum space operators, the generators of space-time translations $P^\mu=(H,\mathbf{P})$ read
\begin{equation}
P^\mu=\int  \frac{d^3\mathbf{p}}{(2\pi)^{3}}p^\mu\sum_{j,s}\bigg\{a^{j s\dagger}_{\mathbf{p}}a^{j s}_{\mathbf{p}}+b^{j s\dagger}_{\mathbf{p}}b^{j s}_{\mathbf{p}}\bigg\}.
\end{equation}
Furthermore, 
defining the vacuum $\ket{0}$ as the state annihilated by $a^{j s}_{\mathbf{p}}$ and $b^{j s}_{\mathbf{p}}$, we can conclude that all states have positive energy.

The spin  of the field is
\begin{equation}
\begin{split}
\mathbf{S}=:-i\int d^3\mathbf{x}\left\{ \dot{\widehat{\psi}}\mathbf{J}\psi-\widehat{\psi}\mathbf{J}\dot{\psi}\right\}:,
\end{split}
\end{equation}
where the components of $\mathbf{J}$ are given by $J^k=\frac{1}{2}\epsilon_{ijk}M^{ij}$, where $M^{\mu\nu}=\frac{i}{4}\left[\gamma^\mu,\gamma^\nu\right]$ are the Lorentz generators of  the $(\tfrac{1}{2},0)\oplus(0,\tfrac{1}{2})$ representation. 
To show that the field contains particles of spin-$1/2$, one can evaluate the action of  the operator $S^3$
on one-particle states with zero momentum in the chiral representation of the Dirac matrices. The resulting relations are
\begin{equation}
S^3a^{j s\dagger}_{\mathbf{0}}\ket{0}= s a^{j s\dagger}_{\mathbf{0}}\ket{0},\quad S^3b^{j s\dagger}_{\mathbf{0}}\ket{0}= s b^{j s\dagger}_{\mathbf{0}}\ket{0}.
\end{equation}

The free theory of second-order fermions in Eq.~(\ref{Lag1}) is invariant under the phase transformation
\begin{equation}
\psi\rightarrow \psi'=e^{i\theta}\psi,\qquad \widehat{\psi}\rightarrow \widehat{\psi}'=\widehat{\psi}e^{-i\theta},
\end{equation}
where $\theta$ is a constant real parameter. The conserved charge associated to this $\mathrm{U}(1)$ global symmetry is
\begin{equation}
\begin{split}
Q=&:i\int d^3\mathbf{x}\bigg\{\widehat{\psi}\dot{\psi}-\dot{\widehat{\psi}}\psi
  \bigg\}:
\\
=&\int \frac{d^3\mathbf{p}}{(2\pi)^{3}}\sum_{j,s}\bigg\{a^{j s\dagger}_{\mathbf{p}}a^{j s}_{\mathbf{p}}-b^{j s\dagger}_{\mathbf{p}}b^{j s}_{\mathbf{p}}\bigg\}.
\end{split}
 \end{equation}
From the commutation relation of this operator with the creation and annihilation operators, one can conclude that  $a^{j s \dagger}_{\mathbf{p}}$ and $b^{j s}_{\mathbf{p}}$ have charge $+1$, whereas $a^{j s }_{\mathbf{p}}$ and $b^{j s \dagger}_{\mathbf{p}}$ have charge $-1$.
Labeling the one-particle states with this eigenvalue, one can show that they are eightfold degenerate
\begin{equation}
\begin{split}
H a^{j s\dagger}_{\mathbf{p}}\ket{0}&\propto H\ket{\mathbf{p},+,j,s}= \omega_{\mathbf{p}}\ket{\mathbf{p},+,j,s},\\
H b^{j s\dagger}_{\mathbf{p}}\ket{0}&\propto H\ket{\mathbf{p},-,j,s}= \omega_{\mathbf{p}}\ket{\mathbf{p},-,j,s}.
\end{split}
\end{equation}

The free theory displays a larger symmetry. Since the field $\psi$ and its dual anticommute $\{\psi_{\alpha}(x),\widehat{\psi}_{\beta}(x)\}=0$, we can write Eq.~(\ref{Lag1}) as
\begin{equation}
\mathcal{L}=\frac{1}{2}\partial^\mu\Psi^T\Omega\partial_\mu\Psi-\frac{m^2}{2}\Psi^T\Omega\Psi,\label{Lag2}
\end{equation}
where $\Psi$ is a column matrix defined as
\begin{equation}
\Psi(x)=\begin{pmatrix}
\widehat{\psi}^T(x)\\
\psi(x)
\end{pmatrix},
\end{equation}
and $\Omega$ is the $8\times 8$ symplectic matrix, written in $4\times 4$ blocks as
\begin{equation}
\Omega=\begin{pmatrix}
0_{4\times4}&1_{4\times4}\\
-1_{4\times4}&0_{4\times4}
\end{pmatrix}.
\end{equation}
Thus, Eq.~(\ref{Lag2}) is symmetric under the global transformations $\Psi\rightarrow \Psi'=S\Psi$ with
$S^T\Omega S=\Omega.$
This is the defining relation for an element of the symplectic group  $\mathrm{Sp}(8,\mathbb{C})$, whose algebra has 36 generators.

The second-order theory is invariant under parity ($P$), charge conjugation ($C$) and time reversal ($T$), and therefore under $CPT$. We define the discrete transformations of the $\psi$ field through their action on the creation operators as follows:
 \begin{gather}
\mathrm{P}a^{js\dagger}_{\mathbf{p}}\mathrm{P}^{-1}=-i(-1)^{j-1} a^{js\dagger}_{-\mathbf{p}},\quad\mathrm{P}b^{js\dagger}_{\mathbf{p}}\mathrm{P}^{-1}=-i(-1)^{j-1}b^{js\dagger}_{-\mathbf{p}},\nonumber\\
\mathrm{C}a^{js\dagger}_{\mathbf{p}}\mathrm{C}^{-1}= b^{js\dagger}_{\mathbf{p}},\quad\quad\quad\mathrm{C}b^{js\dagger}_{\mathbf{p}}\mathrm{C}^{-1}=a^{js\dagger}_{\mathbf{p}},\\
\mathrm{T}a^{js\dagger}_{\mathbf{p}}\mathrm{T}^{-1}=2s a^{j(-s)\dagger}_{-\mathbf{p}},\quad\quad\quad\mathrm{T}b^{is\dagger}_{\mathbf{p}}\mathrm{T}^{-1}=  2s b^{i(-s)\dagger}_{-\mathbf{p}}.\nonumber
\end{gather} 
With this choice, the discrete transformations have the familiar representations
\begin{gather}
\mathrm{P}\psi(x)\mathrm{P}^{-1}=i\gamma^0\psi(\mathcal{P}x),\quad \mathrm{P}\widehat{\psi}(x)\mathrm{P}^{-1}=-i\widehat{\psi}(\mathcal{P}x)\gamma^0,\nonumber\\
\mathrm{C}\psi(x)\mathrm{C}^{-1}=  \mathcal{C} \widehat{\psi}^{\,T},\qquad \mathrm{C}\widehat{\psi}\mathrm{C}^{-1}=\psi^T\mathcal{C},\\
\mathrm{T}\psi(x)\mathrm{T}^{-1}=\mathcal{C}\gamma^5 \psi(\mathcal{T}x),\qquad \mathrm{T}\widehat{\psi}\mathrm{T}^{-1}=-\widehat{\psi}(\mathcal{T}x) \gamma^5\mathcal{C},\nonumber
\end{gather}
where we have defined $\mathcal{P}=\mathrm{diag}(1,-1,-1,-1)$, $\mathcal{T}=\mathrm{diag}(-1,1,1,1)$, and  $\mathcal{C}=-i \gamma^2\gamma^0$ in the chiral representation.
The simplest  $C$, $P$ and $T$  invariant pseudo-Hermitian interactions that can be introduced in this framework are 4-fermion self-interactions represented by a  dimension-four renormalizable operator
\begin{equation}
\begin{split}
\mathcal{L}_{\text{int}}=&   \frac{\lambda_{1}}{2}\left(\widehat{\psi}\psi\right)^{2}
+ \frac{\lambda_{2}}{2}\left( \widehat{\psi}\gamma^{5}\psi\right)  \left(\widehat{\psi}\gamma^{5}\psi\right)\\
&+ \frac{\lambda_{3}}{2}\left( \widehat{\psi}M^{\mu\nu}\psi\right)  \left(\widehat{\psi}M_{\mu\nu}\psi\right).
\label{Lag3}
\end{split}
\end{equation}
Other self-interactions are not independent since they arise in terms of these three through a Fierz transformation. In particular, taking $\lambda_2=\lambda_3=0$, we get the simplest model of self-interacting fermions that at one-loop has vanishing beta function $\beta_{\lambda_1}=0$. The authors of \cite{Vaquera-Araujo:2012jlk} obtained this result in the context of the naive Hermitian second-order theory. All the results contained in \cite{Vaquera-Araujo:2012jlk} and its non-Abelian generalization \cite{Vaquera-Araujo:2013bwa} can be readily applied to the non-Hermitian version of the theory presented here.

Recently, it has been shown in \cite{Durieux:2019siw} that under the helicity-amplitude formalism the mass dimension for an operator with four massless external legs must be six, which may seem to contradict the results in this paper. Nevertheless, as can be seen from the mode expansion, our theory is only valid for $m\neq0$, so it is not possible to have massless fermions with a second-order Lagrangian, leaving room for consistency with the helicity-amplitude formalism.

Summarizing, in this work, we have studied the quantization of the massive field transforming in the $(\tfrac{1}{2},0)\oplus(0,\tfrac{1}{2})$ representation of the RLG that satisfies the Klein-Gordon equation. We have shown that the canonical quantization is free from negative norm states by the redefinition of the field dual and the introduction of a unitary and Hermitian operator $\eta$ that renders the free Lagrangian pseudo-Hermitian according to Eq.(\ref{pseudo-Hermitian}). The resulting canonical quantization of the theory is consistent with microcausality, and the Hamiltonian is bounded from below with real eigenvalues, as expected for a pseudo-Hermitian approach. For consistency, we have verified that the fields have spin-$1/2$, and we have identified the global and discrete symmetries of the theory. Finally, we have sketched the possible interactions for a theory of second-order fermions, including a novel class of renormalizable fermion self-interactions.

To conclude, we remark that this new field would couple with the Higgs field $H$ through a quartic dimensionless coupling
\begin{equation}
\mathcal{L}_{\psi H}=   \frac{\lambda_{H}}{2}\left(\widehat{\psi}\psi\right)H^\dagger H\label{LagHiggs}.
\end{equation}
Furthermore, since $\psi$ has nothing to decay in, it is stable and could play the role of dark matter (DM). From Eq. (\ref{LagHiggs}), the DM, in this case, has a Higgs portal, and the dark matter phenomenology is determined by the mass of the Fermion and its coupling with the Higgs in a similar way as for the scalar dark matter~\cite{Feng:2014vea,Arcadi:2017kky}. The same interaction contributes to the running of the Higgs mass in the same way an extra scalar does, and since the sign of $\lambda_H$ is undetermined, we can choose it in such a way as to cancel the top-quark contribution. This phenomenology and further consequences will be the subject of another article. 
Finally, we express our gratitude to the referee for bringing our attention to Ref. \cite{Sablevice:2023odu}, where the Poincaré algebra is extended to include non-Hermitian generators. In particular, our dual field is compatible with the most general definition identified in \cite{Sablevice:2023odu} that transforms as the dual representation of the full proper Poincaré group.

We thank M. Napsuciale, J. Erler, M. Gorshteyn, and H. Spiesberger for their valuable comments and fruitful discussions. This work was supported by the Grants  No. CONACYT CB-2017-2018/A1-S-13051, No. DGAPA-PAPIIT IN107621, and No. PIIF UNAM. J.O. acknowledges  CONAHCYT national grants, CAV-A is supported by the Consejo nacional de humanidades, ciencias y tecnologías (CONAHCYT) Investigadoras e Investigadores por M\'exico Project No. 749 and SNI 58928. E.P. is grateful for funding from `C\'atedras Marcos Moshinsky' (Fundaci\'on Marcos Moshinsky) and for the support of PASPA-DGAPA, UNAM for a sabbatical leave. R. F. and CAV-A thank  Instituto de F\'{\i}sica de la UNAM for their generous hospitality and support during our stay where this work took place.


\end{document}